\documentclass[journal]{IEEEtran}

\usepackage{graphicx}
\usepackage{amsmath,bbm,epsfig,amssymb,amsfonts,  amstext, verbatim,amsopn,cite,subfigure,multirow,multicol,lipsum}
\usepackage{balance}
\usepackage{url}
\usepackage{amsfonts}
\usepackage{epsfig}
\usepackage{setspace}
\usepackage{stmaryrd}
\usepackage{psfrag}	
\usepackage{multirow}
\usepackage{float}
\usepackage[process=auto]{pstool}
\usepackage{etoolbox}
\usepackage{algorithm}
\usepackage{algorithmic}
\usepackage{lipsum}

\usepackage{pifont}
\allowdisplaybreaks

\newtoggle{Submission}
\togglefalse{Submission}

\iftoggle{Submission}{%

}{%
}

\newcommand*{\Scale}[2][4]{\scalebox{#1}{$#2$}}%

\newtheorem{theo}{Theorem}

\newtheorem{remk}{Remark}

\EndPreamble
\begin{document}

\title{ Cooperative Wireless Backhauling\vspace{-0.2cm}
 }

\author{Vahid Jamali$^\dag$, Nikola Zlatanov$^\ddag$, and Robert Schober$^\dag$ \\
\IEEEauthorblockA{$^\dag$ Friedrich-Alexander University (FAU), Erlangen, Germany \\
 $^\ddag$ Monash University, Melbourne, Australia }\vspace{-0.8cm}
\thanks{This paper is an extended version of the paper which will appear in proceeding of the Int. ITG Conf. Syst., Commun., and Coding 2017.}
}

\maketitle

\begin{abstract}
We consider wireless backhauling for a scenario where two small-cell base stations (SC-BSs) employ the same time and frequency resources for offloading their data to a common macro-cell base station (MC-BS). The two SC-BSs allocate a part of the shared resource to exchange data in order to be able to cooperate to boost the backhaul capacity.  For this scenario, we develop the optimal transmission strategy which, based on the channel state information, determines whether the SC-BSs should exchange data and cooperate or transmit their data independently to the MC-BS. Our numerical results demonstrate the superiority of the proposed cooperative~wireless backhauling protocol compared to  existing protocols in the literature.
\end{abstract}

\section{Introduction} 

Wireless backhauling has recently received considerable attention as a viable, cost-effective, and flexible technology to meet the exponentially increasing data rate demands of future 5G cellular networks \cite{Wireless_Backhaul}. Therefore, developing optimal backhauling protocols which maximize the spectral efficiency is among the most important 5G research challenges. Base station (BS) cooperation and non-orthogonal multiple access (NOMA) are advanced techniques which have been  proposed to improve  spectrum efficiency \cite{NOMA_5G,Coordinate_MIMO}. The latter  refers to transmission schemes where the BSs use the same time and bandwidth resources simultaneously \cite{NOMA_5G} whereas the former specifies the case where the BSs cooperate in order to transmit their data to a shared destination, e.g., a mobile user  or a macro BS. Specifically, cooperation schemes may range from coordinated scheduling and beamforming to full joint data transmission \cite{Coordinate_MIMO}. 

The benefits and challenges of the above advanced techniques for the \textit{radio access network} have been extensively investigated in recent years \cite{NOMA_5G,Coordinate_MIMO,BS_Advance,Coop_Wireless_Backhaul}. Moreover,  many works have optimized radio access protocols assuming finite and constrained backhaul link capacities \cite{BS_Coop_Backhaul,SCell_Lett}. However, exploiting advanced techniques to boost the capacity of the \textit{wireless backhaul network}  has not received the same degree of attention, yet, see \cite{Wireless_Backhaul,BS_Advance,Coop_Wireless_Backhaul}, and the reference therein.

In this paper, we focus on a scenario where two small-cell BSs (SC-BSs) employ the same resources, i.e., non-orthogonal transmission, for offloading their data to a common macro-cell BS (MC-BS). Thereby, \textit{cooperative wireless backhauling} can significantly boost the backhaul capacity via coherent beamforming for data that is available at both SC-BSs, e.g., for the case when the two SC-BSs receive data from the same users. However, in most practical scenarios, different SC-BSs will receive data from different users and hence, have independent information to offload.  Therefore, to exploit the aforementioned advantage of cooperation, additional resources have to be allocated to the exchange of data between the SC-BSs. In this paper, our goal is to determine under what conditions, cooperative wireless backhauling is advantageous despite the extra resources which have to be dedicated to the data exchange between the SC-BSs. To this end, we derive the optimal transmission strategy which,  based on the channel state information (CSI), determines whether the SC-BSs should exchange data and cooperate or transmit their data independently to the MC-BS. Our numerical results reveal that the proposed optimal cooperative wireless backhauling protocol can significantly enhance the backhaul capacity especially when the distance between the SC-BSs~is~small. 

 We note that for the underlying non-orthogonal multiple-access channel assumed in this paper, see Subsection~II-A for a detail description,  an achievable rate region and a statement of the corresponding power allocation problem were given in \cite{AzhangPartI}. Furthermore, the authors in \cite{Davidson} proposed a joint power and bandwidth resource allocation policy for a deterministic non-fading channel based on a simple orthogonal transmission policy in which the bandwidth was divided  into two parts where in each part, one node acts as a relay to assist the other node in order to send its data to the destination. In contrast, in this paper, we derive the optimal non-orthogonal power allocation and resource allocation policies as a function of the instantaneous~CSI for fading channels.  Furthermore, we employ buffers at the user nodes to take advantage of favorable fading condition. 
 In fact, buffer-aided relaying protocols have been derived for different communication scenarios, including rate maximization \cite{NikolaJSAC,ITIEEE,TWCIEEEI}, delay-limited transmission \cite{DelayOptOneWay,TWCIEEEII}, transmission over correlated fading channels \cite{Buffer_Correlated_Fading}, optimal link selection with imperfect CSI \cite{Lett_CSI}, cognitive radio networks \cite{CognitiveAlouini,MostafaLett}, free-space optical (FSO) communications \cite{Marzieh_ICC16,TWC_FSO} and for different network architectures that employ one relay for one-way  \cite{NikolaJSAC,DelayOptOneWay} or two-way  \cite{PopovskiLetter,ITIEEE,TWCIEEEI} transmission and multiple cascaded \cite{HanzoMultiHopCDF,TWC_MultiHop}, parallel \cite{Thompson,NikolaRelaySelection} relays, interference relay channel \cite{MarziehIWCIT}, or diamond relay channel \cite{TWC_Renato}. However, to the best of the authors' knowledge, buffer-aided relay for the cooperative multiple-access channel assumed in this paper has not been investigated, yet.

\textit{Notations:} We use the following notations throughout this paper.  $\mathbbmss{E}\{\cdot\}$ denotes expectation. $|\cdot|$ represents the absolute value of a scalar. Bold small letter $\mathbf{a}=[a_i]$  denotes a vector  with elements $a_i,\,\,\forall i$, and $\mathcal{CN}(0,1)$ denotes a complex Gaussian random variable with zero mean and unit variance.

\section{System Model}\label{SysMod}

In this section, we introduce the considered system model, the adopted transmission scheme, and the required CSI for the proposed~protocol.

\begin{figure}
\centering
\resizebox{0.7\linewidth}{!}{
\pstool[width=1\linewidth]{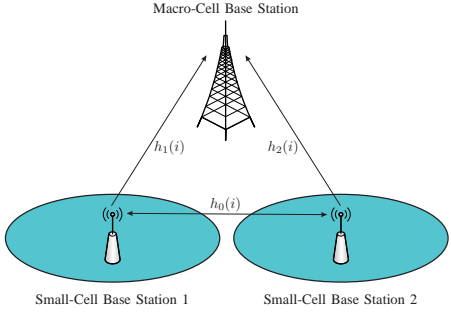}{
\psfrag{U1}[c][c][0.8]{$\text{Small-Cell Base Station~1}$}
\psfrag{U2}[c][c][0.8]{$\text{Small-Cell Base Station~2}$}
\psfrag{D}[c][c][0.8]{$\text{Macro-Cell Base Station}$}
\psfrag{h1D}[c][c][0.8]{$h_{1}(i)$}
\psfrag{h2D}[c][c][0.8]{$h_{2}(i)$}
\psfrag{h12}[c][c][0.8]{$h_{0}(i)$}
}}
\caption{Illustration of BS cooperation for data offloading of two SC-BSs to a common MC-BS.}
\label{FigSysMod}
\vspace{-0.4cm}
\end{figure}

\subsection{System Model}

We consider a wireless backhauling network consisting of two SC-BSs and an MC-BS where the SC-BSs cooperate to send their data to the common MC-BS, see Fig. \ref{FigSysMod}. This   communication setup can be used to model the backhauling networks for the following practical scenarios: \textit{i)} Small-cell networks where the mobile nodes in a building floor send their  data  to  an SC-BS on the roof of the building and  the  SC-BS  forwards the  information  to  the MC-BS. Thereby, the SC-BSs of neighboring buildings can employ cooperative wireless backhauling to enhance the backhaul capacity. \textit{ii)} Communication in trains where the users in each wagon send their data to an SC-BS on the roof of the train wagon and  neighboring SC-BSs cooperatively send their data to a nearby infrastructure MC-BS.

\begin{figure}
\centering
\resizebox{0.8\linewidth}{!}{
\pstool[width=0.9\linewidth]{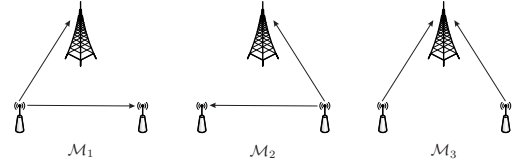}{
\psfrag{A}[c][c][0.7]{$\mathcal{M}_1$}
\psfrag{B}[c][c][0.7]{$\mathcal{M}_2$}
\psfrag{C}[c][c][0.7]{$\mathcal{M}_3$}
}}\vspace{-0.2cm}
\caption{The three possible transmission modes for the considered BS cooperation scheme where $\mathcal{M}_1$ and $\mathcal{M}_2$ allow the data exchange between the SC-BSs and $\mathcal{M}_3$ employs cooperative transmission of SC-BSs to the MC-BS.}
\label{FigMod3}
\vspace{-0.4cm}
\end{figure}

We assume that the SC-BSs employ the same time and frequency resources for offloading their data to the MC-BS. Time is assumed to be divided into slots of equal length indexed by $i=1,\dots,N$,  and each node transmits codewords which span one time
slot. We also assume
that all communication links are impaired
by additive white Gaussian noise (AWGN) and block fading, i.e., the channel
coefficients are constant during one time slot and change from
one time slot to the next. We assume half-duplex transmission because of its simplicity and feasibility\footnote{Full-duplex nodes have been reported in the literature \cite{DuarteFD}. However, they entail high hardware complexity for efficient self-interference suppression. Hence, in this paper, we focus on half-duplex communication.}, where the SC-BSs either transmit or receive. Based on these assumptions, three transmission modes  are possible for the network, which are denoted by $\mathcal{M}_k,\,\,k=1,2,3$, see Fig. \ref{FigMod3}. The received codewords for each transmission mode can be modelled~as
\begin{IEEEeqnarray}{lll}
  \mathcal{M}_1:\quad &  Y_2(i) &= h_{0}(i)X_1(i)+Z_2(i) \label{Gaussian1}\IEEEyesnumber \IEEEyessubnumber \\
  &  Y_m(i) &= h_{1}(i)X_1(i) +Z_m(i), \IEEEyessubnumber \\
   \mathcal{M}_2:\quad & Y_1(i) &= h_{0}(i)X_2(i)+Z_1(i)  \label{Gaussian2}\IEEEyesnumber\IEEEyessubnumber \\
 &  Y_m(i) &=  h_{2}(i)X_2(i)+Z_m(i), \IEEEyessubnumber \\
  \mathcal{M}_3:\quad &  Y_m(i) &= h_{1}(i)X_1(i) + h_{2}(i)X_2(i)+Z_m(i), \label{Gaussian3}\IEEEyesnumber
\end{IEEEeqnarray}
where $X_j(i)$, $j\in\{1,2\}$, $Y_j(i),$ $j\in\{1,2,m\}$,  and $Z_j(i)$,  $j\in\{1,2,m\}$, denote the transmitted codeword of  node $j$, the  received
codeword at node $j$, and the noise at node $j$ in the $i$-th time slot, respectively. For $X_j(i)$, $Y_j(i)$, and $Z_j(i)$, superscripts $j=1, 2$, and $m$ are used to denote SC-BS 1, SC-BS2, and  MC-BS, respectively.   We assume that the noises at the nodes are independent from each other and from the transmitted codewords. Moreover, we assume the noise variance at all receivers is given by $[\sigma_n^2]_{\mathrm{dB}}=WN_0+N_F$ where $W$, $N_0$, and $N_F$ denote the channel bandwidth, the noise power spectral density (in  dB/Hz),  and  the  noise figure  (in  dB)  of  the  receiver, respectively. Furthermore, $h_{0}(i)$, $h_{1}(i)$, and $h_{2}(i)$ denote the complex-valued channel coefficients of the links between SC-BS 1 and SC-BS 2, SC-BS 1 and the MC-BS, and SC-BS 2 and the MC-BS in the $i$-th time slot, respectively.  The squares  of the channel coefficient amplitudes in the $i$-th time slot are denoted by $s_l(i)=|h_l(i)|^2,\,\,l\in\{0,1,2\}$. Moreover, we introduce sets $\mathcal{S}_{l}$, which contain the possible fading states $\mathbf{s}=(s_{0},s_{1},s_{2})\in\mathcal{S}_{0}\times\mathcal{S}_{1}\times\mathcal{S}_{2}$\footnote{In this paper, we drop time index $i$ for fading state  $\mathbf{s}$  for notational simplicity.}.  The $s_l,\,\,\forall l$,  are assumed to be mutually independent, ergodic, and stationary random processes. Furthermore, 
we assume that the fading states have continuous probability density functions denoted by $f_l(s_l)$, $\forall l$. Since the noise is AWGN, the SC-BSs transmit Gaussian distributed codewords to maximize their data rates, i.e., $X_j(i)$ is comprised of symbols which are zero-mean rotationally invariant complex Gaussian random variables with variance $P_j, \, j\in\{1,2\}$. Hence, $P_j$ represents the transmit power of SC-BS $j$, which is assumed to be fixed for all time slots.  Additionally, we define $\gamma_j(i)=P_j/\sigma_n^2,\,\,j\in\{1,2\}$, as the transmit signal-to-noise ratio (SNR) of SC-BS $j$. We also use the definition $C(x)\triangleq W\log_2(1+x)$ for notational~simplicity.

\subsection{Transmission Scheme}

Let $B_1$ and $B_2$ denote two infinite-size buffers at SC-BS~1 and SC-BS 2, respectively. Moreover, $Q_j(i), \,\, j\in\{1,2\}$, denotes the amount of normalized information in bits/symbol available in buffer $B_j$ at the end of the $i$-th time slot. The coding scheme which will be presented in the following is in principal a modification of the coding scheme developed for ideal full-duplex communication  in \cite{AzhangPartI} to the case of half-duplex communication which is assumed in this paper. Recall that our goal is to investigate, based on the CSI of the involved links, when the SC-BSs should employ cooperation and when they should transmit their data independently to the MC-BS. To this end, we assume that the SC-BSs employ rate splitting between two types of messages: \textit{i)} a message which is intended for decoding at the MC-BS without BS cooperation (non-cooperative message); and \textit{ii)} a message which is intended for decoding at the MC-BS via BS cooperation (cooperative message). In the following, the corresponding coding schemes, transmission rates, and dynamics of the queues at the buffers for the three transmission modes are presented.

\noindent
\textbf{Transmission mode $\mathcal{M}_1$:}  SC-BS 1 broadcasts codeword $X_1(i)$ to SC-BS 2 and the MC-BS. SC-BS 2 receives $Y_2(i)$ according to (\ref{Gaussian1}a) and the MC-BS receives $Y_m(i)$ according to~(\ref{Gaussian1}b).

\textit{Encoding:}  For this mode, the codeword of SC-BS 1 is constructed as follows
\begin{IEEEeqnarray}{lll}\label{CodewordX1}
  X_1(i)=\sqrt{\alpha_1^{(1)}(\mathbf{s}) P_1} U_1(i) + \sqrt{\alpha_2^{(1)}(\mathbf{s})P_1} V_1(i) 
\end{IEEEeqnarray}
where $U_1(i)\sim \mathcal{CN}(0,1)$ is an auxiliary  Gaussian codeword carrying the information of the cooperative message at rate $R_1^{\mathrm{c}}(\mathbf{s})$ bits/symbol to be decoded at SC-BS 2 in the $i$-th time slot and to be decoded at the MC-BS in some future time slots via BS cooperation. In contrast, $V_1(i)\sim \mathcal{CN}(0,1)$ is an  Gaussian codeword which carries the information of the non-cooperative message at rate $R_1^{\mathrm{nc}}(\mathbf{s})$ bits/symbol intended for directly decoding  at the MC-BS without cooperation. Moreover, $\alpha_1^{(1)}(\mathbf{s})$ and $\alpha_2^{(1)}(\mathbf{s})$ are the fractions of power $P_1$ allocated to codewords $U_1(i)$ and $V_1(i)$ for mode $\mathcal{M}_1$, respectively,  where $\alpha_j^{(1)}(\mathbf{s})\in[0,1],\,\,j=1,2$ and $\alpha_1^{(1)}(\mathbf{s})+\alpha_2^{(1)}(\mathbf{s})=1$ have to hold. Furthermore, for the construction of $U_1(i)\sim \mathcal{CN}(0,1)$, SC-BS 1 employs the superposition coding introduced in \cite{AzhangPartI}. In particular, the codebook for $U_1(i)$ is divided into $2^{n(R_1^{\mathrm{c}}(\mathbf{s})-T_1(\mathbf{s}))}$ uniform disjoint partitions where $T_1(\mathbf{s})$ is an auxiliary variable satisfying $T_1(\mathbf{s})<R_1^{\mathrm{c}}(\mathbf{s})$. These partitions are indexed by $J_1(\mathbf{s})\in\{1,\dots,2^{n(R_1^{\mathrm{c}}(\mathbf{s})-T_1(\mathbf{s}))}\}$ where each partition is comprised of $2^{nT_1(\mathbf{s})}$ codewords\footnote{We assume that the SC-BS's codebooks and the mapping of the SC-BS's codewords to different partitions are known at all nodes.}. 

 \textit{Decoding:}  SC-BS 2 decodes $U_1(i)$ and treats $V_1(i)$ as noise. Then, it determines to which partition (of the codebook for $U_1(i)$) the decoded codeword belongs and stores only the index of the partition,  $J_1(\mathbf{s})$, in its buffer $B_2$. The MC-BS does not perform  decoding in the $i$-th time slot, instead it stores the received codeword $Y_m(i)$, and waits until SC-BS 2 sends the partition index $J_1(\mathbf{s})$. After the MC-BS has received the partition index from SC-BS 2 in some future time slots,  it first decodes codeword $U_1(i)$ from the received codeword $Y_m(i)$ by treating $V_1(i)$ as noise and searching only among the codewords in the partition with index $J_1(\mathbf{s})$  in the codebook for $U_1(i)$. For successful transmission in this mode, the transmission rate of  SC-BS 1 must satisfy
\begin{IEEEeqnarray}{rll}\label{RateM1}
  R_1^{\mathrm{c}}(\mathbf{s})\,&\leq C_{12}^{\mathrm{c}}(\mathbf{s}) \IEEEeqnarraynumspace\IEEEyesnumber\IEEEyessubnumber \\
  R_1^{\mathrm{nc}}(\mathbf{s})\,&\leq C_{1m}^{\mathrm{nc}}(\mathbf{s}) \IEEEyessubnumber \\
  T_1(\mathbf{s})\,&\leq \min\{R_1^{\mathrm{c}}(\mathbf{s}),C_{1m}^{\mathrm{c}}(\mathbf{s})\}, \IEEEyessubnumber 
\end{IEEEeqnarray}
where $C_{12}^{\mathrm{c}}(\mathbf{s})=C\big(\alpha_1^{(1)}(\mathbf{s}) \gamma_1 s_0/(1+\alpha_2^{(1)}(\mathbf{s}) \gamma_1 s_0)\big)$, $C_{1m}^{\mathrm{c}}(\mathbf{s})=C\big( \alpha_1^{(1)}(\mathbf{s}) \gamma_1 s_1/(1+ \alpha_2^{(1)}(\mathbf{s}) \gamma_1 s_1) \big)$, and $C_{1m}^{\mathrm{nc}}(\mathbf{s})=C(\alpha_2^{(1)}(\mathbf{s}) \gamma_1 s_1)$.  

 \textit{Dynamics of the Queues:} After SC-BS 2 has received the messages transmitted by SC-BS 1 in the $i$-th time
slot,  the  amount  of  information  in  buffer   $B_2$  increases  to $Q_2(i)=Q_2(i-1)+R^c_1(\mathbf{s})-T_1(\mathbf{s})$. Moreover, the  amount  of  information  in  buffer   $B_1$  does not change, i.e., $Q_1(i)=Q_1(i-1)$.

\noindent
\textbf{Transmission mode $\mathcal{M}_2$:} The coding scheme for mode $\mathcal{M}_2$ is identical to the one  for mode $\mathcal{M}_1$ with SC-BSs 1 and 2 switching roles. Hence, in order to avoid repetition, we provide only the results  for mode $\mathcal{M}_2$ which we require in the remainder of the paper and do not state the coding strategy in detail.  Let $\beta_1^{(2)}(\mathbf{s})$ and $\beta_2^{(2)}(\mathbf{s})$  denote the fractions of power $P_2$ allocated to the cooperative codewords $U_2(i)$ and the non-cooperative codewords $V_2(i)$ for mode $\mathcal{M}_2$, respectively,  where $\beta_j^{(2)}(\mathbf{s})\in[0,1],\,\,j=1,2$ and $\beta_1^{(2)}(\mathbf{s})+\beta_2^{(2)}(\mathbf{s})= 1$ have to hold. Moreover, for  successful decoding in this mode, the transmission rate of  SC-BS 2 must satisfy
\begin{IEEEeqnarray}{rll}\label{RateM2}
  R_2^{\mathrm{c}}(\mathbf{s})\,&\leq C_{21}^{\mathrm{c}}(\mathbf{s}) \IEEEeqnarraynumspace\IEEEyesnumber\IEEEyessubnumber \\
  R_2^{\mathrm{nc}}(\mathbf{s})\,&\leq C_{2m}^{\mathrm{nc}}(\mathbf{s}) \IEEEyessubnumber \\
  T_2(\mathbf{s}) \,&\leq \min\{R_2^{\mathrm{c}}(\mathbf{s}),C_{2m}^{\mathrm{c}}(\mathbf{s})\}, \IEEEyessubnumber 
\end{IEEEeqnarray}
where $C_{21}^{\mathrm{c}}(\mathbf{s})=C\big(\beta_1^{(2)}(\mathbf{s}) \gamma_2 s_0/(1+\beta_2^{(2)}(\mathbf{s}) \gamma_2 s_0)\big)$, $C_{2m}^{\mathrm{c}}(\mathbf{s})=C\big( \beta_1^{(2)}(\mathbf{s}) \gamma_2 s_2/(1+ \beta_2^{(2)}(\mathbf{s}) \gamma_2 s_2) \big)$, and $C_{2m}^{\mathrm{nc}}(\mathbf{s})=C(\beta_2^{(2)}(\mathbf{s}) \gamma_2 s_2)$.    After SC-BS 1 has received the messages transmitted by SC-BS 2 in the $i$-th time slot,  the  amount  of  information  in  buffer  $B_1$  increases  to $Q_1(i)=Q_1(i-1)+R^c_2(\mathbf{s})-T_2(\mathbf{s})$ and the  amount  of  information  in  buffer   $B_2$  does not change, i.e., $Q_2(i)=Q_2(i-1)$.

\noindent
\textbf{Transmission mode $\mathcal{M}_3$:}  SC-BSs 1 and 2 simultaneously transmit codewords $X_1(i)$ and $X_2(i)$ to the MC-BS, respectively, and the MC-BS receives $Y_m(i)$ according to (\ref{Gaussian3}).

\textit{Encoding:}  For this mode, the codwords of the SC-BSs are constructed as 
\begin{IEEEeqnarray}{lll}\label{CodewordX1X2}
  X_1(i)=\sqrt{\alpha_1^{(3)}(\mathbf{s}) P_1} U_1(i) &+ \sqrt{\alpha_2^{(3)}(\mathbf{s})P_1} U_2(i) \nonumber \\ &  + \sqrt{\alpha_3^{(3)}(\mathbf{s})P_1} V_1(i) \IEEEyesnumber\IEEEyessubnumber \\
  X_2(i)= \sqrt{\beta_1^{(3)}(\mathbf{s}) P_2} U_2(i)&+ \sqrt{\beta_1^{(3)}(\mathbf{s})P_2} U_1(i)  \nonumber \\ & + \sqrt{\beta_3^{(3)}(\mathbf{s})P_2} V_2(i), \IEEEyessubnumber 
\end{IEEEeqnarray}
where $U_j(i)\sim \mathcal{CN}(0,1), \,\,j=1,2$, is a Gaussian codeword carrying the refinement information at rate $R_j^{\mathrm{c}}(\mathbf{s})$ bits/symbol to be used by the MC-BS to decode the codewords which have been transmitted to the MC-BS in former time slots and is intended to be decoded via BS cooperation\footnote{Note that SC-BS 1 constructs $U_1(i)$ by extracting the refinement information from buffer $B_1$ while SC-BS 2 can construct $U_1(i)$ since it is generated by its own message. In a similar manner, both SC-BSs can construct $U_2(i)$.}. In contrast, $V_j(i)\sim \mathcal{CN}(0,1), \,\,j=1,2$, is a  Gaussian codeword carrying the information of the non-cooperative message at rate $R_j^{\mathrm{nc}}(\mathbf{s})$ bits/symbol intended to be decoded at the MC-BS without cooperation. Moreover, $\alpha_1^{(3)}(\mathbf{s})$, $\alpha_2^{(3)}(\mathbf{s})$, and $\alpha_3^{(3)}(\mathbf{s})$ are the fractions of power $P_1$  allocated to codewords $U_1(i)$, $U_2(i)$, and $V_1(i)$, respectively, where $\alpha_j^{(3)}(\mathbf{s})\in[0,1],\,\,j=1,2,3$ and $\sum_{j=1}^{3}\alpha_j^{(3)}(\mathbf{s})=1$ have to hold. Similarly,  $\beta_1^{(3)}(\mathbf{s})$, $\beta_2^{(3)}(\mathbf{s})$, and $\beta_3^{(3)}(\mathbf{s})$ are the fractions of power $P_2$ which are allocated to codewords $U_2(i)$, $U_1(i)$, and $V_2(i)$, respectively,  where $\beta_j^{(3)}(\mathbf{s})\in[0,1],\,\,j=1,2,3$ and $\sum_{j=1}^{3}\beta_j^{(3)}(\mathbf{s})=1$ have~to~hold.

\textit{Decoding:}
The MC-BS employs successive decoding \cite{Cover}. In particular, the MC-BS first decodes   $U_j(i), \,\,j=1,2$, and treats  $V_j(i), \,\,j=1,2$,  as noise. Then, it subtracts the contribution of  $U_j(i), \,\,j=1,2$, from the received codeword $Y_m(i)$ and decodes $V_j(i), \,\,j=1,2$. The transmission rates of the SC-BSs in each time slot are limited by the capacity region of the multiple-access channel with correlated sources \cite{Cover,AzhangPartI}, and the amount of information stored in buffers $B_1$ and $B_2$. Therefore, for successful transmission in this mode, the transmission rates of the SC-BSs  must satisfy 
\begin{IEEEeqnarray}{lll}\label{RateM3}
  R_j^{\mathrm{nc}}(\mathbf{s}) \leq C_{jr}^{\mathrm{nc}}(\mathbf{s}),\qquad j=1,2 \IEEEeqnarraynumspace\IEEEyesnumber\IEEEyessubnumber \\
  R_j^{\mathrm{c}}(\mathbf{s}) \leq Q_j(i-1),\qquad j=1,2 \IEEEyessubnumber \\
  R_1^{\mathrm{nc}}(\mathbf{s})+R_2^{\mathrm{nc}}(\mathbf{s})\leq C_{\mathrm{sum}}^{\mathrm{nc}}(\mathbf{s})  \IEEEyessubnumber \\
  R_1^{\mathrm{nc}}(\mathbf{s})+R_2^{\mathrm{nc}}(\mathbf{s})+R_1^{\mathrm{c}}(\mathbf{s})+R_2^{\mathrm{c}}(\mathbf{s})\leq C_{\mathrm{sum}}^{\mathrm{c}}(\mathbf{s}) \IEEEyessubnumber 
\end{IEEEeqnarray}
where $C_{1r}^{\mathrm{nc}}(\mathbf{s})=C(\alpha_3^{(3)}(\mathbf{s}) \gamma_1s_1)$, $C_{2r}^{\mathrm{nc}}(\mathbf{s})=C(\beta_3^{(3)}(\mathbf{s}) \gamma_2 s_2)$, $C_{\mathrm{sum}}^{\mathrm{nc}}(\mathbf{s}) =C(\alpha_3^{(3)}(\mathbf{s}) \gamma_1s_1+\beta_3^{(3)}(\mathbf{s}) \gamma_2 s_2)$, and $C_{\mathrm{sum}}^{\mathrm{c}}(\mathbf{s}) = C\Big(\gamma_1s_1+\gamma_2s_2+2\sqrt{(\alpha_1^{(3)}(\mathbf{s})+\alpha_2^{(3)}(\mathbf{s}))(\beta_1^{(3)}(\mathbf{s})+\beta_2^{(3)}(\mathbf{s})) \gamma_1 \gamma_2 s_1 s_2} \Big)$.

\textit{Dynamics of the Queues:} After the transmission in the $i$-th time slot, the amounts of information in buffers $B_1$ and
$B_2$ have decreased to $Q_1(i)=Q_1(i-1)- R_1^{\mathrm{c}}(\mathbf{s})$ and $Q_2(i)=Q_2(i-1)- R_2^{\mathrm{c}}(\mathbf{s})$, respectively.

\subsection{CSI Requirements}

Throughout this paper, we assume that the MC-BS has
full knowledge of the CSI of all links and is responsible for determining which  transmission  mode  is
selected in each time slot and for conveying the transmission strategy to the SC-BSs, cf.   Theorem \ref{OptProt}. Moreover, we assume that the channel states change slow enough such that the signaling overhead caused by channel estimation and feedback  is negligible compared to the amount of transmitted information.

\section{Backhaul Capacity Maximization}

In this section, we formulate the backhaul capacity maximization problem and solve it to obtain the optimal protocol.

\subsection{Problem Formulation}

Let $\bar{R}_{\mathrm{sum}}^{\mathrm{c}}$ and $\bar{R}_{\mathrm{sum}}^{\mathrm{nc}}$ denote the average sum rates of SC-BS~1 and SC-BS 2 with and without BS cooperation, respectively. In this paper, our goal is to optimally choose the aforementioned transmission modes in each time slot based on the CSI of the involved links such that the backhaul capacity (sum rate of the SC-BSs), $\bar{R}_{\mathrm{sum}}^{\mathrm{c}}+\bar{R}_{\mathrm{sum}}^{\mathrm{nc}}$, is maximized. To this end, we introduce binary variables $q_{k}(\mathbf{s})\in\{0,1\},\,\,k\in\{1,2,3\}$, where $q_k(\mathbf{s})=1$ if transmission mode $\mathcal{M}_k$ is selected in the $i$-th time slot and $q_k(\mathbf{s})=0$ if it is not selected. Moreover, since in each time slot only one of the transmission modes can be selected, only one of the mode selection variables is equal to one and the others are zero, i.e., $\sum_{k=1}^{3}q_k(\mathbf{s})=1,\,\,\forall \mathbf{s}$~holds.  For notational convenience of the problem formulation, we define the following average capacity rates:  $\bar{C}_{12}^{\mathrm{c}}=\mathbbmss{E}\{q_1(\mathbf{s})C_{12}^{\mathrm{c}}(\mathbf{s})\}$, $\bar{C}_{1m}^{\mathrm{nc}}=\mathbbmss{E}\{q_1(\mathbf{s}) C_{1m}^{\mathrm{nc}}(\mathbf{s})\}$, $\bar{C}_{21}^{\mathrm{c}}=\mathbbmss{E}\{q_2(\mathbf{s})C_{21}^{\mathrm{c}}(\mathbf{s})\}$, $\bar{C}_{2m}^{\mathrm{c}}=\mathbbmss{E}\{q_2(\mathbf{s}) C_{2m}^{\mathrm{c}}(\mathbf{s})\}$, $\bar{C}_{\mathrm{sum}}^{\mathrm{c}}=\mathbbmss{E}\{q_3(\mathbf{s})C_{\mathrm{sum}}^{\mathrm{c}}(\mathbf{s})\}$, and $\bar{C}_{\mathrm{sum}}^{\mathrm{nc}}=\mathbbmss{E}\{q_3(\mathbf{s}) C_{\mathrm{sum}}^{\mathrm{nc}}(\mathbf{s})\}$.

We note that since each SC-BS knows the information in the other SC-BS's buffer (because it is its own message), both queues can be viewed as a single \textit{virtual queue} containing the common message that both SC-BSs know and thus, they can cooperatively and coherently transmit it to the MC-BS. Thereby, for the virtual queue to be stable \cite{Neely}, the average arrival rate at the virtual queue, i.e.,   $\bar{C}^{\mathrm{c}}_{21}-\bar{C}_{2m}^{\mathrm{c}}$ for buffer $B_1$ plus $\bar{C}_{12}^{\mathrm{c}}-\bar{C}_{1m}^{\mathrm{c}}$ for buffer $B_2$, should be \textit{equal to or less than} the average departure rate of the virtual queue, i.e., $\bar{C}_{\mathrm{sum}}^{\mathrm{c}}-\bar{C}_{\mathrm{sum}}^{\mathrm{nc}}$ for both buffers. Before proceeding
further, we highlight and exploit the following useful results from \cite[Lemma 2]{ITIEEE}. We note that although these results were originally developed for the bidirectional relay channel, they are also valid for the system model considered in this paper. Hence, in order to avoid repetition,  we do not reprove them for the system model in this paper. In particular, it is shown in \cite[Lemma 2]{ITIEEE} that for the optimal throughput-maximizing policy, while the queue is rate-stable, the gap between the average arrival and departure rates of the queue has to vanish.   In this case, the effect of the number of time slots in which the virtual queue  does not have enough information to supply (or equivalently \textit{none} of the buffers has enough information to supply), due to constraint (\ref{RateM3}b), becomes negligible as $N\to\infty$. Using these results,  the backhaul capacity optimization problem for the considered system can be formulated as
\begin{IEEEeqnarray}{rll}\label{ProbOpt}
    {\underset{\mathbf{q}\in\boldsymbol{\mathcal{Q}},\boldsymbol{\alpha}\in\boldsymbol{\mathcal{A}},\boldsymbol{\beta}\in\boldsymbol{\mathcal{B}}}{\mathrm{maximize}}}\,\,\, & \bar{C}_{\mathrm{sum}}^{\mathrm{nc}} + \bar{C}_{1m}^{\mathrm{nc}}+\bar{C}_{2m}^{\mathrm{nc}} + \bar{C}_{12}^{\mathrm{c}} + \bar{C}_{21}^{\mathrm{c}} \nonumber\\
     \mathrm{subject\,\, to} \,\,\, & \bar{C}_{\mathrm{sum}}^{\mathrm{nc}}+\bar{C}_{12}^{\mathrm{c}} + \bar{C}_{21}^{\mathrm{c}} = \bar{C}_{\mathrm{sum}}^{\mathrm{c}}+\bar{C}_{1m}^{\mathrm{c}}+\bar{C}_{2m}^{\mathrm{c}}, \qquad
\end{IEEEeqnarray}
where $\mathbf{q}=[q_k(\mathbf{s})],\,\,\forall \mathbf{s},k$, is the mode selection variable with feasible set $\boldsymbol{\mathcal{Q}}=\{\mathbf{q}|q_{k}(\mathbf{s})\in\{0,1\},\,\,\forall \mathbf{s},k,\,\,\wedge\,\, \sum_k q_k(\mathbf{s})=1,\,\,\forall \mathbf{s}\}$. Moreover, $\boldsymbol{\alpha}=\big[\alpha_j^{(k)}(\mathbf{s})\big],\,\,\forall \mathbf{s},k,j$, collects the power sharing variables for SC-BS 1 and its  feasible set is given by $\boldsymbol{\mathcal{A}}=\{\boldsymbol{\alpha}|\alpha_j^{(k)}(\mathbf{s})\in[0,1]\,\,\wedge\,\,\sum_j \alpha_j^{(k)}(\mathbf{s})=1,\,\,\forall \mathbf{s},k\}$. Similarly, $\boldsymbol{\beta}=\big[\beta_j^{(k)}(\mathbf{s})\big],\,\,\forall \mathbf{s},k,j$, collects the power sharing variables for SC-BS 2 and its feasible set is given by $\boldsymbol{\mathcal{B}}=\{\boldsymbol{\beta}|\beta_j^{(k)}(\mathbf{s})\in[0,1]\,\,\wedge\,\,\sum_j \beta_j^{(k)}(\mathbf{s})=1,\,\,\forall \mathbf{s},k\}$. Furthermore, the constraint in (\ref{ProbOpt}) is the optimal queue condition developed in \cite[Lemma 2]{ITIEEE}.
\begin{figure*}
\begin{IEEEeqnarray}{lll}\label{MutualSet}
\Scale[0.8]{
\mathcal{K}_1  =  \bigg\{  \mathbf{s}\big |    \min\{\gamma_1s_1,\gamma_2s_2\}  \geq \left(  \mu-\frac{1}{2}  \right)  \left(1+\gamma_1s_1+\gamma_2s_2 \right)\bigg\} }   \,\,  \qquad  \IEEEyesnumber\IEEEyessubnumber \\
\Scale[0.8]{
\mathcal{K}_2 = \bigg\{  \mathbf{s}\big |    \min\left\{\sqrt{\frac{\gamma_2s_2}{\gamma_1s_1}},\sqrt{\frac{\gamma_1s_1}{\gamma_2s_2}}\right\}  \geq   \frac{1-\mu}{\mu} \left[ 1+\left(\sqrt{\gamma_1s_1}+\sqrt{\gamma_2s_2}\right)^2 \right] \bigg\}}  \,\,  \qquad \IEEEyessubnumber \\
\Scale[0.8]{
\mathcal{K}_3  = \bigg\{  \mathbf{s}\bigg | \sqrt{\frac{\gamma_2s_2}{\gamma_1s_1}} < \frac{1-\mu}{\mu} \left[ 1+\left(\sqrt{\gamma_1s_1}+\sqrt{\gamma_2s_2}\right)^2 \right] \,\, \wedge \,\, \gamma_2s_2 < \left(\mu-\frac{1}{2}\right)\left(1+\gamma_1s_1+\gamma_2s_2 \right) \bigg\}}  \IEEEyessubnumber  \\
\Scale[0.8]{
\mathcal{K}_4  = \bigg\{  \mathbf{s}\bigg | \sqrt{\frac{\gamma_1s_1}{\gamma_2s_2}} < \frac{1-\mu}{\mu} \left[ 1+\left(\sqrt{\gamma_1s_1}+\sqrt{\gamma_2s_2}\right)^2 \right] \,\, \wedge \,\, \gamma_1s_1 \leq \left(\mu-\frac{1}{2}\right)\left(1+\gamma_1s_1+\gamma_2s_2 \right) \bigg\}.}  \quad\quad  \IEEEyessubnumber
\end{IEEEeqnarray}
\hrulefill
\end{figure*}

\subsection{Optimal Backhauling Protocol}

\begin{figure}
\centering
\resizebox{0.8\linewidth}{!}{
\pstool[width=1\linewidth]{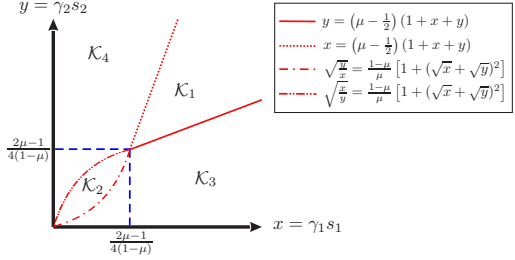}{
\psfrag{X}[l][c][0.9]{$x=\gamma_1s_1$}
\psfrag{Y}[c][c][0.9]{$y=\gamma_2s_2$}
\psfrag{K1}[c][c][0.9]{$\mathcal{K}_1$}
\psfrag{K2}[c][c][0.9]{$\mathcal{K}_2$}
\psfrag{K3}[c][c][0.9]{$\mathcal{K}_3$}
\psfrag{K4}[c][c][0.9]{$\mathcal{K}_4$}
\psfrag{L}[c][c][0.8]{$\frac{2\mu-1}{4(1-\mu)}$}
\psfrag{T1}[l][c][0.7]{$y=\left(\mu-\frac{1}{2}\right)\left(1+x+y\right)$}
\psfrag{T2}[l][c][0.7]{$x=\left(\mu-\frac{1}{2}\right)\left(1+x+y\right)$}
\psfrag{T3}[l][c][0.7]{$\sqrt{\frac{y}{x}}=\frac{1-\mu}{\mu}\left[1+(\sqrt{x}+\sqrt{y})^2\right]$}
\psfrag{T4}[l][c][0.7]{$\sqrt{\frac{x}{y}}=\frac{1-\mu}{\mu}\left[1+(\sqrt{x}+\sqrt{y})^2\right]$}
}}
\vspace{-0.03cm}
\caption{Four mutually exclusive fading regions, i.e., $\mathcal{K}_1$, $\mathcal{K}_2$, $\mathcal{K}_3$, and $\mathcal{K}_4$, required for specification of the optimal values of the power sharing variables in transmission mode $\mathcal{M}_3$.}
\label{FigPowerReg}
\vspace{-0.3cm}
\end{figure}

Before formally stating the optimal protocol as the solution of  (\ref{ProbOpt}), we introduce some auxiliary variables which we require for the statement of the protocol.   First,  in the optimal protocol, the instantaneous link capacities  are weighted by a constant $\mu$ which we refer to as selection weight\footnote{The selection weight $\mu$ is in fact the Lagrange multiplier corresponding to the constraint in (\ref{ProbOpt}).}. 
The value of $\mu$ depends on the channel statistics and can be obtained offline and used as long as the channel statistics remain unchanged. Second, we define sets $\mathcal{K}_1$, $\mathcal{K}_2$, $\mathcal{K}_3$, and $\mathcal{K}_4$ based on the fading states as given in (\ref{MutualSet}) at the top of the next page.
These four mutually exclusive fading sets are illustrated in Fig. \ref{FigPowerReg}. The optimal power sharing for transmission mode $\mathcal{M}_3$ depends on the set to which the fading state belongs. Third, we give the optimal power sharing policy in terms of the following variables
\begin{IEEEeqnarray}{lll}\label{PowerSimp}
\alpha^{(1)}(\mathbf{s}) &\triangleq \alpha^{(1)}_1(\mathbf{s}) = 1- \alpha^{(1)}_2(\mathbf{s}) \IEEEyesnumber\IEEEyessubnumber  \\
\beta^{(2)}(\mathbf{s}) &\triangleq \beta^{(2)}_1(\mathbf{s}) = 1- \beta^{(2)}_2(\mathbf{s})   \IEEEyessubnumber  \\
\alpha^{(3)}(\mathbf{s}) &\triangleq \alpha^{(3)}_1(\mathbf{s})+\alpha^{(3)}_2(\mathbf{s}) = 1- \alpha^{(3)}_3(\mathbf{s}) \IEEEyessubnumber  \\
\beta^{(3)}(\mathbf{s}) &\triangleq \beta^{(3)}_1(\mathbf{s})+\beta^{(3)}_2(\mathbf{s}) = 1- \beta^{(3)}_3(\mathbf{s}). \IEEEyessubnumber
\end{IEEEeqnarray}
Moreover, for given $\alpha^{(3)}(\mathbf{s})$ and $\beta^{(3)}(\mathbf{s})$, the optimal power sharing policy reveals that sharing the power between auxiliary codewords $U_1(i)$ and $U_2(i)$ at SC-BS 1 and SC-BS 2 does not change the sum rate as long as at least one of the buffers can supply enough information. In other words, there exists a degree of freedom in choosing $\alpha^{(3)}_1(\mathbf{s})$,  $\alpha^{(3)}_2(\mathbf{s})$, $\beta^{(3)}_1(\mathbf{s})$, and $\beta^{(3)}_2(\mathbf{s})$  as long as  $\alpha^{(3)}_1(\mathbf{s})+\alpha^{(3)}_2(\mathbf{s})=\alpha^{(3)}(\mathbf{s})$ and  $\beta^{(3)}_1(\mathbf{s})+\beta^{(3)}_2(\mathbf{s})=\beta^{(3)}(\mathbf{s})$ hold.

\begin{theo}[Optimal Backhauling Protocol]\label{OptProt}
The optimal mode selection and power sharing policies which maximize the capacity of the considered wireless backhauling network with BS cooperation are given in the following. The optimal mode selection policy is given by
\begin{IEEEeqnarray}{lll}\label{OptPolicy}
q_{k^*}(i)=
   \begin{cases}
     1, &k^*= {\underset{k=1,2,3}{\arg \, \max}} \,\, \Lambda_k(i)  \\
     0, &\mathrm{otherwise}
\end{cases}
 \IEEEyesnumber
\end{IEEEeqnarray}
where $\Lambda_k(\mathbf{s})$ is referred to as the selection metric and  given~by
\begin{IEEEeqnarray}{lll}\label{SelecMet}
    \Lambda_1(\mathbf{s}) = (1-\mu) C_{12}^{\mathrm{c}}(\mathbf{s}) + \mu C_{1m}^{\mathrm{c}}(\mathbf{s}) \IEEEyesnumber\IEEEyessubnumber  \\
    \Lambda_2(\mathbf{s}) =  (1-\mu) C_{21}^{\mathrm{c}}(\mathbf{s}) + \mu C_{2m}^{\mathrm{c}}(\mathbf{s})  \IEEEyessubnumber  \\
    \Lambda_3(\mathbf{s}) = (1-\mu) C_{\mathrm{sum}}^{\mathrm{nc}}(\mathbf{s}) + \mu C_{\mathrm{sum}}^{\mathrm{c}}(\mathbf{s}) . \IEEEyessubnumber
\end{IEEEeqnarray}
Whereas,  the optimal power sharing policy is given by
\begin{IEEEeqnarray}{lll}\label{OptPower}
\alpha^{(1)}(\mathbf{s}) &= \beta^{(2)}(\mathbf{s}) = 1 \IEEEyesnumber\IEEEyessubnumber  \\
\alpha^{(3)}(\mathbf{s}) &= \begin{cases}
1, &\mathrm{if}\,\, \mathbf{s} \in \mathcal{K}_2 \cup \mathcal{K}_4 \\
\left[ \frac{-b_1+\sqrt{b_1^2+4a_1c_1}}{2a_1}  \right]^2, &\mathrm{if}\,\, \mathbf{s} \in \mathcal{K}_3 \\
 \left(\mu-\frac{1}{2}\right)\frac{1+\gamma_1s_1+\gamma_2s_2}{\gamma_1s_1}, &\mathrm{if}\,\, \mathbf{s} \in \mathcal{K}_1
\end{cases} \IEEEyessubnumber  \\
\beta^{(3)}(\mathbf{s}) &= \begin{cases}
1, &\mathrm{if}\,\, \mathbf{s} \in \mathcal{K}_2 \cup \mathcal{K}_3  \\
\left[ \frac{-b_2+\sqrt{b_2^2+4a_2c_2}}{2a_2}  \right]^2, &\mathrm{if}\,\, \mathbf{s} \in \mathcal{K}_4 \\
 \left(\mu-\frac{1}{2}\right)\frac{1+\gamma_1s_1+\gamma_2s_2}{\gamma_2s_2}, &\mathrm{if}\,\, \mathbf{s} \in \mathcal{K}_1
\end{cases} \IEEEyessubnumber
\end{IEEEeqnarray}
where $a_1=\frac{2-\mu}{\mu}\gamma_1s_1$, $b_1=\frac{1-\mu}{\mu}\sqrt{\frac{\gamma_1s_1}{\gamma_2s_2}}(1+\gamma_1s_1+\gamma_2s_2)$, $c_1=1+\gamma_1s_1$, $a_2=\frac{2-\mu}{\mu}\gamma_2s_2$, $b_2=\frac{1-\mu}{\mu}\sqrt{\frac{\gamma_2s_2}{\gamma_1s_1}}(1+\gamma_1s_1+\gamma_2s_2)$, and $c_2=1+\gamma_2s_2$. 
Furthermore, $\mu\in(\frac{1}{2},1)$ is a constant which is obtained numerically by solving the following equation
\begin{IEEEeqnarray}{lll}\label{EqConst}
 \iiint\limits_{\mathbf{s}} \Big[  q_1(\mathbf{s})\big(C_{12}^{\mathrm{c}}(\mathbf{s})-C_{1m}^{\mathrm{c}}(\mathbf{s})\big) + q_2(\mathbf{s})\big(C_{21}^{\mathrm{c}}(\mathbf{s})-C_{2m}^{\mathrm{c}}(\mathbf{s})\big) \nonumber \\ + q_3(\mathbf{s}) \big(C_{\mathrm{sum}}^{\mathrm{nc}}(\mathbf{s}) - C_{\mathrm{sum}}^{\mathrm{c}}(\mathbf{s}) \big) \Big]  f_0(s_0)f_1(s_1)f_2(s_2)  \mathrm{d}\mathbf{s}=0, \,\, \quad
\end{IEEEeqnarray}
where $q_k(\mathbf{s})$ and $[\alpha^{(k)}(\mathbf{s}),\beta^{(k)}(\mathbf{s})]$ in the above equation have to be substituted from (\ref{OptPolicy}) and (\ref{OptPower}), respectively. 
\end{theo}


\iftoggle{Submission}{%
Due to space constraints, we give the proof in \cite{ITG2017_arXiv} which is an extended version of this paper. In the following remarks, we highlight some insights that the protocol in Theorem~\ref{OptProt} provides.
}{%
  \begin{IEEEproof}
Please refer to the Appendix.  
\end{IEEEproof}  } 

\begin{remk}
The mode selection metric $\Lambda_k(i)$ introduced in (\ref{SelecMet}) is a weighted sum of the capacity terms
in each time slot where the weight, $\mu$, is constant. In each time slot, the mode with the highest value of the selection metric is selected. Since the fading states have continuous probability density functions, the probability that $\Lambda_k(i)=\Lambda_{k'}(i),\,\,k\neq k'$, holds is zero. Hence, the  selection  policy  in  (\ref{OptPolicy})  indicates  that,  for  any  fading  state  $\mathbf{s}=(s_0,s_1,s_2)$,  the  choice  of  the  optimal
transmission  mode  is  unique.  In  other  words,  for  a  given  fading  state,  it  is  sub-optimal  to  share  the
resources between the transmission modes and only one of the transmission modes should be used.
Hence, \textit{adaptive mode selection} is the key to maximize the capacity of the considered wireless backhauling protocol with BS cooperation.
\end{remk}

\begin{remk}\label{RemkPowerReg}
The following observations can be made from the optimal protocol in Theorem \ref{OptProt}. \textit{i)} Rate splitting between the cooperative and non-cooperative messages for  modes $\mathcal{M}_1$ and $\mathcal{M}_2$, $U_j(i)$ and $V_j(i)$, $j=1,2$, is strictly sub-optimal, cf. (\ref{OptPower}a). \textit{ii)} Rate splitting between the cooperative messages for  mode $\mathcal{M}_3$, $U_1(i)$ and $U_2(i)$, does not change/improve the backhaul capacity, cf. (\ref{PowerSimp}). \textit{iii)} Rate splitting between the cooperative and non-cooperative messages for transmission mode $\mathcal{M}_3$, $U_j(i)$ and $V_j(i)$, $j=1,2$, can improve the backhaul capacity for certain fading states depending on to which set $\mathcal{K}_l,\,\,l=1,2,3,4$,  fading state $\mathbf{s}$ belongs, cf. (\ref{OptPower}b) and~(\ref{OptPower}c).
\end{remk}

\begin{remk}
We note that the advantages of data exchange between the SC-BSs and data buffering at the SC-BSs  come at the expense of an increased end-to-end delay. However, with some modifications to the optimal protocol, the average delay can be bounded (using e.g. a similar technique as in \cite[Subsection~IV-B]{ITIEEE} for bidirectional relaying) which causes only a small loss in the achievable backhaul capacity. However, a delay analysis of the proposed protocol is beyond the scope of the current work and is left for future research.
\end{remk}

\section{Numerical Results}

In this section, we first present several benchmark schemes. Subsequently, we evaluate the performance of the proposed protocol with respect to the benchmark schemes.

\subsection{Benchmark Scheme}

We consider the following three benchmark schemes:

\textit{Orthogonal Transmission without BS Cooperation:} SC-BSs 1 and 2 transmit their data to the MC-BS in odd and even time slots, respectively. The average capacity of this backhauling protocol is given by $\tau^{\mathrm{o,nc}}=0.5\mathbbmss{E}\{C(\gamma_1s_1)+C(\gamma_2s_2)\}$.

\textit{Non-Orthogonal Transmission without BS Cooperation:} SC-BSs 1 and 2 transmit simultaneously their independent data to the MC-BS. The MC-BS employs successive decoding to recover  the SC-BSs' data \cite{Cover}. The average capacity of this backhauling protocol is given by $\tau^{\mathrm{no,nc}}=\mathbbmss{E}\{C(\gamma_1s_1+\gamma_2s_2)\}$.

\textit{Non-Orthogonal Genie-Aided BS Cooperation:} As performance upper bound, we consider the case when the SC-BSs have identical information (without spending any resources for data exchange) to send to the MC-BS. The SC-BSs perform coherent data transmission such that their signals add up coherently at the MC-BS \cite{AzhangPartI}.  The average capacity of this idealistic backhauling scheme is given by $\tau^{\mathrm{no,c}}=\mathbbmss{E}\{C(\gamma_1s_1+\gamma_2s_2+2\sqrt{\gamma_1\gamma_2s_1s_2})\}$.

By comparing our proposed protocol with the above benchmark schemes, we are able to determine whether the origin of the backhaul capacity improvement is the non-orthogonal transmission and/or the BS~cooperation.  Note that the protocol in \cite{AzhangPartI} was developed for ideal full-duplex nodes. Moreover, the half-duplex protocol proposed in \cite[Section~V]{Davidson} was given as the solution to an optimization problem, i.e., not in closed form, which is valid only for a specific SNR range, i.e., $\mathbbmss{E}\{\gamma_1s_0\}\geq \mathbbmss{E}\{\gamma_1s_1\}$ and $\mathbbmss{E}\{\gamma_2s_0\}\geq \mathbbmss{E}\{\gamma_2s_2\}$. In contrast, our goal in this paper is to determine under what conditions, e.g., \textit{in which SNR range},  \textit{half-duplex cooperation} is beneficial. Hence, we cannot use the protocols in \cite{AzhangPartI} and \cite{Davidson} as benchmarks.

\subsection{Performance Evaluation}

\begin{table}
\label{Table:Parameter}
\caption{Values of the Numerical Parameters \cite{Wireless_Backhaul,Fast_Train}.\vspace{-0.2cm}} 
\begin{center}
\scalebox{0.6}
{
\begin{tabular}{|| c | c  | c | c ||}
  \hline
   Symbol & Value & Symbol & Value \\ \hline \hline
 $(d_0,d_1,d_2)$ &  $(100,1000,1000)$ m &
   $d^{\mathrm{ref}}$ &  $80$ m \\ \hline
   $(P_1,P_2)$ & $(200,200)$ mW ($(23,23)$ dBm) & $W$ & $20$ MHz \\ \hline
  $N_0$ & $-114$ dBm/MHz & 
  $N_F$ &  $5$ dB  \\ \hline 
   $\lambda$  & $85.7$ mm ($3.5$ GHz) &
    $\nu$ &  $3.5$ \\ \hline           
    $(G^{\mathrm{Tx}}_1, G^{\mathrm{Tx}}_2)$ & $(5,5)$ dBi & $(G^{\mathrm{Rx}}_1, G^{\mathrm{Rx}}_2,G^{\mathrm{Rx}}_m)$ & $(5,5,10)$ dBi \\ \hline
\end{tabular}
}
\end{center}
\vspace{-0.5cm}
\end{table}

In this subsection, we numerically evaluate the performance of the proposed protocol for the considered wireless backhauling network with BS cooperation in Rayleigh fading for $N=10^6$ fading blocks. We assume a distance-dependent path-loss model given by $\mathbbmss{E}\{s_l\}= \Big[\frac{\lambda\sqrt{G^{\mathrm{Tx}}_j G^{\mathrm{Rx}}_j }}{4\pi d^{\mathrm{ref}}}\Big]^2 \times \left[\frac{d^{\mathrm{ref}}}{d_l}\right]^\nu$ where $\lambda$ is the wavelength of the signal, $G^{\mathrm{Tx}}_j$ and $G^{\mathrm{Rx}}_j$ are the antenna gains of node $j\in\{1,2,m\}$ for transmitting and receiving, respectively, $d^{\mathrm{ref}}$ is a reference distance for the antenna far-field, $d_l$ is the distance between the transmitter and the receiver, and $\nu$ is the path-loss exponent. The default values of the system parameters used in the numerical results are given in Table~I.

\begin{figure}
\centering
\resizebox{0.85\linewidth}{!}{
\psfragfig{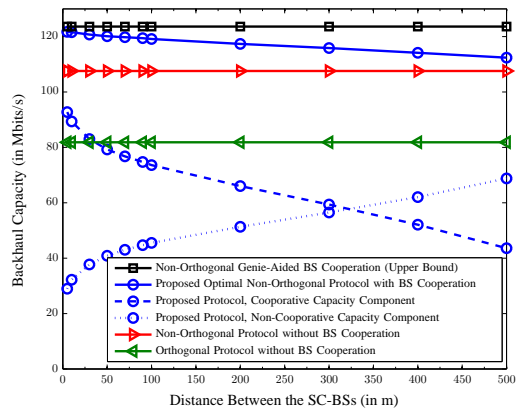}}
\vspace{-0.5cm}
\caption{Backhaul capacity (in Mbits/s) versus the distance between the SC-BSs (in m).}
\label{Fig:Rsum_Distance}
\vspace{-0.3cm}
\end{figure}

In Fig. \ref{Fig:Rsum_Distance}, we show the backhaul capacity (in Mbits/s) versus the distance between the SC-BSs, $d_0$, (in m).  In addition to the results for the proposed protocol and the considered benchmark schemes,  in Fig.~\ref{Fig:Rsum_Distance}, we include the cooperative component,  $\bar{R}_{\mathrm{sum}}^{\mathrm{c}}$, and the non-cooperative component, $\bar{R}_{\mathrm{sum}}^{\mathrm{nc}}$, of the backhaul capacity achieved by the proposed protocol. From Fig.~\ref{Fig:Rsum_Distance}, we observe that as the distance between the SC-BSs decreases, the cooperative component of the capacity increases and the non-cooperative component decreases, i.e., the SC-BSs share more data to enable cooperative transmission to the MC-BS. Moreover, from high to low values of $d_0$, the backhaul capacity achieved by the proposed protocol increases from the backhaul capacity achieved by  non-orthogonal transmission without BS cooperation to the upper bound of non-orthogonal genie-aided BS cooperation. Furthermore, the orthogonal protocol is outperformed by all non-orthogonal protocols by a large~margin.

\begin{figure}
\centering
\resizebox{0.85\linewidth}{!}{
\psfragfig{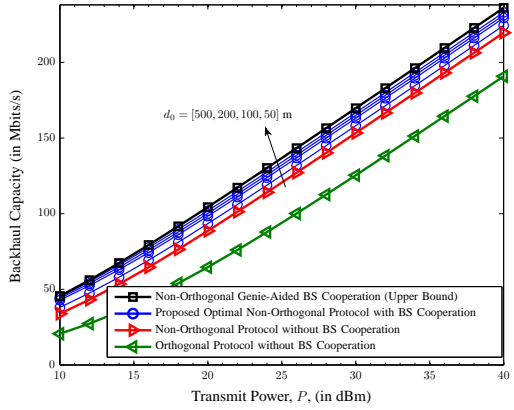}}
\vspace{-0.5cm}
\caption{Backhaul capacity (in Mbits/s) versus SC-BS transmit powers $P_1=P_2=P$ (in dBm) for $d_0=[500,200,100,50]$ m.  For the proposed protocol, $d_0$ decreases along the direction of the arrow.}
\label{Fig:Rsum_SNR}
\vspace{-0.3cm}
\end{figure}

In Fig.~\ref{Fig:Rsum_SNR}, we show the backhaul capacity (in Mbits/s) versus the SC-BS transmit powers $P_1=P_2=P$ (in dBm) for $d_0=[500,200,100,50]$ m. As expected, the backhaul capacity increases as the transmit power increases. Moreover, we see from Fig.~\ref{Fig:Rsum_SNR} that for $d_0 = 50$ m, the backhaul capacity achieved by the proposed protocol is very close to that of the upper bound for the considered range of the SC-BS transmit power. In fact, a distance of $50$ m is a typical distance between wagons of a train or neighboring houses in residential areas.  Therefore, backhauling of the small-cell networks deployed in trains or in residential areas can be potential applications of the proposed cooperative wireless backhaul protocol.

\section{Conclusions}

We studied a wireless backhauling scenario where two SC-BSs use the same time and frequency resources to cooperatively send their data to a common MC-BS.
We derived the optimal transmission strategy which, based on the CSI, determines whether the SC-BSs should exchange their data and cooperate or transmit their data independently to the MC-BS.   Our  numerical  results  showed that the proposed optimal cooperative wireless backhaul  protocol can significantly enhance the backhaul capacity especially when the distance between the SC-BSs is small.

\iftoggle{Submission}{%

}{%

\appendix
 
In this appendix, we solve the optimization problem given in (\ref{ProbOpt}). In particular, we first relax the binary constraints $q_k(\mathbf{s})\in\{0, 1\}$ to $0\leq q_k(\mathbf{s})\leq 1$. Then, we solve the relaxed problem and  show that one of the solutions always lies at the boundaries of $0\leq q_k(\mathbf{s})\leq 1$. Thus, this solution of the relaxed problem solves the original problem as well.

In the following, we investigate the
Karush-Kuhn-Tucker (KKT) necessary conditions \cite{Boyd} for the problem in (\ref{ProbOpt})  and show that these
necessary conditions result in a unique value for the backhaul capacity. Denoting the Lagrange multiplier corresponding to the constraint in (\ref{ProbOpt}) by $\mu$, the Lagrangian function for the optimization problem in (\ref{ProbOpt}) is obtained as
\begin{IEEEeqnarray}{ll}\label{LagFunction}
   \mathcal{L}(\mathbf{q},\boldsymbol{\alpha},\boldsymbol{\beta},\mu)  = \bar{C}_{\mathrm{sum}}^{\mathrm{nc}} + \bar{C}_{1m}^{\mathrm{nc}}+\bar{C}_{2m}^{\mathrm{nc}} + \bar{C}_{12}^{\mathrm{c}} + \bar{C}_{21}^{\mathrm{c}}   \nonumber \\ \qquad  + \mu \big( \bar{C}_{\mathrm{sum}}^{\mathrm{c}}+\bar{C}_{1m}^{\mathrm{c}}+\bar{C}_{2m}^{\mathrm{c}} - \bar{C}_{\mathrm{sum}}^{\mathrm{nc}} - \bar{C}_{12}^{\mathrm{c}} - \bar{C}_{21}^{\mathrm{c}} \big).\quad 
\end{IEEEeqnarray}
The optimal mode selection and power sharing variables for a given $\mu$ are obtained by calculating the derivatives of the Lagrangian function with respect to $\mathbf{q}$, $\boldsymbol{\alpha}$, and $\boldsymbol{\beta}$, respectively, cf. (\ref{OptPolicy}) and (\ref{OptPower}). Then, we substitute the optimal $\mathbf{q}$, $\boldsymbol{\alpha}$, and $\boldsymbol{\beta}$ as a function of $\mu$ in  constraint (\ref{ProbOpt}) and find $\mu$ such that this constraint holds, cf. (\ref{EqConst}). 

\subsection{Optimal Mode Selection Variables}

The derivatives of Lagrangian function in (\ref{LagFunction}) with respect to $q_k(\mathbf{s})$ are given by
\begin{IEEEeqnarray}{rll}\label{StationaryMode}
    \frac{\partial\mathcal{L}}{\partial q_1(\mathbf{s})} \,\,&= \Pr\{\mathbf{s}\} \big[ (1-\mu) C_{12}^{\mathrm{c}}(\mathbf{s}) + \mu C_{1m}^{\mathrm{c}}(\mathbf{s}) + C_{1m}^{\mathrm{nc}}(\mathbf{s}) \big] \qquad \IEEEyesnumber \IEEEyessubnumber \\
    \frac{\partial\mathcal{L}}{\partial q_2(\mathbf{s})} \,\,&= \Pr\{\mathbf{s}\} \big[ (1-\mu) C_{21}^{\mathrm{c}}(\mathbf{s}) + \mu C_{2m}^{\mathrm{c}}(\mathbf{s}) + C_{2m}^{\mathrm{nc}}(\mathbf{s}) \big] \quad  \IEEEyessubnumber \\
    \frac{\partial\mathcal{L}}{\partial q_3(\mathbf{s})} \,\,&= \Pr\{\mathbf{s}\}  \big[ (1-\mu) C_{\mathrm{sum}}^{\mathrm{nc}}(\mathbf{s}) + \mu C_{\mathrm{sum}}^{\mathrm{c}}(\mathbf{s})  \big]. \quad  \IEEEyessubnumber 
\end{IEEEeqnarray}
As we show in remainder of this appendix, the optimal $\mu$ has to be in the interval $(\frac{1}{2},1)$ in order for constraint (\ref{ProbOpt}) to hold. Hence, for a given channel state, the derivative $\frac{\partial\mathcal{L}}{\partial q_k(\mathbf{s})}$ is always positive. Moreover, for ergodic fading with continuous probability density function, the probability that $\Lambda_k(\mathbf{s}) = \Lambda_{k'}(\mathbf{s}),\,\,k\neq k'$, holds is zero where $\Lambda_k(\mathbf{s}) = \frac{\partial\mathcal{L}}{\partial q_k(\mathbf{s})}$. Therefore, since $\sum_{k=1}^{3}q_k(\mathbf{s})=1,\,\,\forall \mathbf{s}$ has to hold, we select the optimal mode corresponding to the largest value of $\Lambda_k(\mathbf{s})$. This is equivalent to the optimal mode selection policy given in (\ref{OptPolicy}). Note that the terms $C_{1m}^{\mathrm{nc}}(\mathbf{s})$ ad $C_{2m}^{\mathrm{nc}}(\mathbf{s})$ have been dropped in $\Lambda_1(\mathbf{s})$ and $\Lambda_2(\mathbf{s})$ in (\ref{SelecMet}) since, in the following,  we prove  that $\alpha_2^{(1)}(\mathbf{s})=0$ and $\beta_2^{(1)}(\mathbf{s})=0$ have to hold if $q_1(\mathbf{s})=1$ and $q_2(\mathbf{s})=1$, respectively.

\subsection{Optimal Power Sharing Variables}

Since $q_k(\mathbf{s})$ is either zero or one, we only have to obtain the optimal $\alpha_j^{(k)}(\mathbf{s})$ and $\beta_j^{(k)}(\mathbf{s})$ if $q^*_k(\mathbf{s})=1$ holds. Assuming $q_1(\mathbf{s})=1$, we calculate the derivatives of  the Lagrangian function in (\ref{LagFunction}) with respect to $\alpha_1^{(1)}(\mathbf{s})$ and $\alpha_2^{(1)}(\mathbf{s})$.   This leads~to
\begin{IEEEeqnarray}{rll}\label{StationaryPower1}
    \frac{\partial\mathcal{L}}{\partial \alpha_1^{(1)}(\mathbf{s})} &= 0 \qquad \IEEEyesnumber \IEEEyessubnumber  \\
     \frac{\partial\mathcal{L}}{\partial \alpha_2^{(1)}(\mathbf{s})} &= \frac{\Pr\{\mathbf{s}\}}{\ln 2} \frac{\gamma_1(1-\mu) (s_1-s_0) }{(1+\alpha_2^{(k)}(\mathbf{s})\gamma_1s_1)(1+\alpha_2^{(k)}(\mathbf{s})\gamma_1s_0)}  \qquad \IEEEyessubnumber
\end{IEEEeqnarray}
Moreover, for ergodic fading with continuous probability density function, we obtain $\Pr\{s_1=s_0\}=0$. Hence, the derivative $\frac{\partial\mathcal{L}}{\partial \alpha_2^{(1)}(\mathbf{s})}$ is either positive or negative. This leads to
\begin{IEEEeqnarray}{rll}\label{PowerSharM1}
    [\alpha_1^{(1)}(\mathbf{s}), \alpha_2^{(1)}(\mathbf{s})] = \begin{cases}
    [1,0],\quad & \mathrm{if} \,\,s_0>s_1 \\
    [0,1], & \mathrm{if} \,\, s_0 < s_1 
    \end{cases}
\end{IEEEeqnarray}
However, if $s_0 < s_1$ holds, transmission mode $\mathcal{M}_1$ cannot be selected. In particular, assuming $s_0 < s_1$, we obtain $\Lambda_1(\mathbf{s})=C(\gamma_1s_1)$  by substituting the optimal power sharing variables in (\ref{PowerSharM1}) into (\ref{SelecMet}a). Now, we can replace mode $\mathcal{M}_1$ with mode $\mathcal{M}_3$ assuming $\alpha_3^{(3)}(\mathbf{s})=1$ and $\beta_3^{(3)}(\mathbf{s})=1$ and improve the backhaul capacity. Hence, for $s_0 < s_1$, mode $\mathcal{M}_1$ cannot be selected in the optimal mode selection policy since the resulting achievable rate  is sub-optimal. In a similar manner, we can show that $\beta_1^{(1)}(\mathbf{s})=1$ and $\beta_2^{(1)}(\mathbf{s})=0$ have to hold if $q_2(\mathbf{s})=1$.

Assuming $q_3(\mathbf{s})=1$, we calculate the derivatives of  the Lagrangian function in (\ref{LagFunction}) with respect to $\alpha_j^{(3)}(\mathbf{s})$ and $\beta_j^{(3)}(\mathbf{s}),\,\,j=1,2,3$.   This leads to
\begin{IEEEeqnarray}{rll}\label{StationaryPower3}
    \frac{\partial\mathcal{L}}{\partial \alpha_1^{(3)}(\mathbf{s})} &=\frac{\partial\mathcal{L}}{\partial \alpha_2^{(3)}(\mathbf{s})}  \nonumber \\ & = \frac{\Pr\{\mathbf{s}\}}{\ln 2\,\, d} \frac{\mu(\beta_1^{(3)}(\mathbf{s})+\beta_2^{(3)}(\mathbf{s})) \gamma_1 \gamma_2 s_1 s_2}{1+\gamma_1s_1+\gamma_2s_2+2d}\qquad \IEEEyesnumber \IEEEyessubnumber  \\ 
        \frac{\partial\mathcal{L}}{\partial \alpha_3^{(3)}(\mathbf{s})} &= \frac{\Pr\{\mathbf{s}\}}{\ln 2} \frac{(1-\mu)\gamma_1s_1}{1+\alpha_3^{(3)}(\mathbf{s}) \gamma_1s_1+\beta_3^{(3)}(\mathbf{s}) \gamma_2 s_2} \IEEEyessubnumber \\        
         \frac{\partial\mathcal{L}}{\partial \beta_1^{(3)}(\mathbf{s})} &=\frac{\partial\mathcal{L}}{\partial \beta_2^{(3)}(\mathbf{s})} \nonumber \\ &  = \frac{\Pr\{\mathbf{s}\}}{\ln 2\,\, d} \frac{\mu(\alpha_1^{(3)}(\mathbf{s})+\alpha_2^{(3)}(\mathbf{s})) \gamma_1 \gamma_2 s_1 s_2}{1+\gamma_1s_1+\gamma_2s_2+2d}\qquad \IEEEyessubnumber \\
        \frac{\partial\mathcal{L}}{\partial \beta_3^{(3)}(\mathbf{s})} &= \frac{\Pr\{\mathbf{s}\}}{\ln 2} \frac{(1-\mu)\gamma_2s_2}{1+\alpha_3^{(3)}(\mathbf{s}) \gamma_1s_1+\beta_3^{(3)}(\mathbf{s}) \gamma_2 s_2} \IEEEyessubnumber 
\end{IEEEeqnarray}
where $d=\sqrt{(\alpha_1^{(3)}(\mathbf{s})+\alpha_2^{(3)}(\mathbf{s}))(\beta_1^{(3)}(\mathbf{s})+\beta_2^{(3)}(\mathbf{s})) \gamma_1 \gamma_2 s_1 s_2}$. Since $\frac{\partial\mathcal{L}}{\partial \alpha_1^{(3)}(\mathbf{s})} =\frac{\partial\mathcal{L}}{\partial \alpha_2^{(3)}(\mathbf{s})}$ and $\frac{\partial\mathcal{L}}{\partial \beta_1^{(3)}(\mathbf{s})} =\frac{\partial\mathcal{L}}{\partial \beta_2^{(3)}(\mathbf{s})}$ hold, we can conclude that for given $\alpha_1^{(3)}(\mathbf{s})+\alpha_2^{(3)}(\mathbf{s})$ and $\frac{\partial\mathcal{L}}{\partial \beta_1^{(3)}(\mathbf{s})} + \frac{\partial\mathcal{L}}{\partial \beta_2^{(3)}(\mathbf{s})}$, sharing the power between auxiliary codewords $U_1(i)$ and $U_2(i)$ at SC-BS 1 and SC-BS 2 does not change the sum rate.  For clarity of the rest of the analysis, we define $\alpha \triangleq \alpha_1^{(3)}(\mathbf{s})+\alpha_2^{(3)}(\mathbf{s})$ and $\bar{\alpha} \triangleq \alpha_3^{(3)}(\mathbf{s})$, $\beta \triangleq \beta_1^{(3)}(\mathbf{s})+\beta_2^{(3)}(\mathbf{s})$, and $\bar{\beta} \triangleq \beta_3^{(3)}(\mathbf{s})$. Note that $\frac{\partial\mathcal{L}}{\partial \alpha}=\frac{\partial\mathcal{L}}{\partial \alpha_1^{(3)}(\mathbf{s})} =\frac{\partial\mathcal{L}}{\partial \alpha_2^{(3)}(\mathbf{s})}$ and $\frac{\partial\mathcal{L}}{\partial \beta} =\frac{\partial\mathcal{L}}{\partial \beta_1^{(3)}(\mathbf{s})} =\frac{\partial\mathcal{L}}{\partial \beta_2^{(3)}(\mathbf{s})}$ hold. In the following, we consider nine possible mutually exclusive cases for the relations  $\frac{\partial\mathcal{L}}{\partial \alpha } \gtreqless \frac{\partial\mathcal{L}}{\partial \bar{\alpha} }$ and $\frac{\partial\mathcal{L}}{\partial\beta } \gtreqless \frac{\partial\mathcal{L}}{\partial \bar{\beta} }$ and find  the necessary condition for optimality in each case based on the fading gains.

\textit{Case 1:} If we assume that $\frac{\partial\mathcal{L}}{\partial \alpha } < \frac{\partial\mathcal{L}}{\partial \bar{\alpha} }$ and $\frac{\partial\mathcal{L}}{\partial\beta } < \frac{\partial\mathcal{L}}{\partial \bar{\beta} }$ hold, we obtain $\alpha=0$ and $\beta=0$. For this case, we have to consider the limiting case when $\alpha,\beta \to 0$. Substituting these values in (\ref{StationaryPower3}), we obtain the necessary condition for $\frac{\partial\mathcal{L}}{\partial \alpha } < \frac{\partial\mathcal{L}}{\partial \bar{\alpha} }$ and $\frac{\partial\mathcal{L}}{\partial\beta } < \frac{\partial\mathcal{L}}{\partial \bar{\beta} }$ as follows
\begin{IEEEeqnarray}{rll}\label{NecEquCase1}
    \sqrt{\frac{\gamma_2s_2}{\gamma_1s_1}} \leq \frac{1-\mu}{\mu} \underset{\alpha,\beta \to 0}{\lim} \sqrt{\frac{\alpha}{\beta}} \IEEEyesnumber \IEEEyessubnumber  \\ 
    \sqrt{\frac{\gamma_1s_1}{\gamma_2s_2}} \leq \frac{1-\mu}{\mu} \underset{\alpha,\beta \to 0}{\lim} \sqrt{\frac{\beta}{\alpha}}, \IEEEyessubnumber 
\end{IEEEeqnarray}
respectively. Note that regarding how $\alpha$ and $\beta$ approach zero,  the set of fading states $\mathbf{s}$ which satisfies both aforementioned conditions can be non-empty only if both right hand sides of (\ref{NecEquCase1}a) and (\ref{NecEquCase1}b) are larger than one which leads to $\mu<\frac{1}{2}$. However, we will show in Subsection C of this appendix that  $\mu>\frac{1}{2}$ has to hold for the optimal protocol. Hence, we conclude that, for the optimal solution, $\frac{\partial\mathcal{L}}{\partial \alpha } < \frac{\partial\mathcal{L}}{\partial \bar{\alpha} }$ and $\frac{\partial\mathcal{L}}{\partial\beta } < \frac{\partial\mathcal{L}}{\partial \bar{\beta} }$ cannot hold for any fading state.

\textit{Case 2:} If we assume that $\frac{\partial\mathcal{L}}{\partial \alpha } > \frac{\partial\mathcal{L}}{\partial \bar{\alpha} }$ and $\frac{\partial\mathcal{L}}{\partial\beta } < \frac{\partial\mathcal{L}}{\partial \bar{\beta} }$ hold, we obtain $\alpha=1$ and $\beta=0$. Substituting these values into (\ref{StationaryPower3}), we obtain the necessary condition for $\frac{\partial\mathcal{L}}{\partial \alpha } > \frac{\partial\mathcal{L}}{\partial \bar{\alpha} }$ as 
\begin{IEEEeqnarray}{rll}
    \frac{(1-\mu)\gamma_1s_1}{1+\gamma_2s_2} \leq 0
\end{IEEEeqnarray}
which occurs with probability zero, i.e., $\Pr\{s_1\leq 0\}= 0$, considering that $\mu\in(\frac{1}{2},1)$ holds. Therefore, we conclude that $\frac{\partial\mathcal{L}}{\partial \alpha } > \frac{\partial\mathcal{L}}{\partial \bar{\alpha} }$ and $\frac{\partial\mathcal{L}}{\partial\beta } < \frac{\partial\mathcal{L}}{\partial \bar{\beta} }$ cannot hold  for the optimal solution.

\textit{Case 3:} If we assume that $\frac{\partial\mathcal{L}}{\partial \alpha } < \frac{\partial\mathcal{L}}{\partial \bar{\alpha} }$ and $\frac{\partial\mathcal{L}}{\partial\beta } > \frac{\partial\mathcal{L}}{\partial \bar{\beta} }$ hold, we obtain $\alpha=0$ and $\beta=1$. Similar to the reasoning given for Case 3,  we can conclude that the necessary condition for the optimality of this case cannot hold for any fading state.

\textit{Case 4:} If we assume that $\frac{\partial\mathcal{L}}{\partial \alpha } = \frac{\partial\mathcal{L}}{\partial \bar{\alpha} }$ and $\frac{\partial\mathcal{L}}{\partial\beta } < \frac{\partial\mathcal{L}}{\partial \bar{\beta} }$ hold, we obtain $\beta=0$. Substituting this value in (\ref{StationaryPower3}), we obtain the necessary condition for $\frac{\partial\mathcal{L}}{\partial \alpha } = \frac{\partial\mathcal{L}}{\partial \bar{\alpha} }$ as 
\begin{IEEEeqnarray}{rll}
    \frac{(1-\mu)\gamma_1s_1}{1+\bar{\alpha}\gamma_1s_2+\gamma_2s_2} = 0
\end{IEEEeqnarray}
which occurs with probability zero, i.e., $\Pr\{s_1=0\}=0$. Therefore, we conclude that $\frac{\partial\mathcal{L}}{\partial \alpha } = \frac{\partial\mathcal{L}}{\partial \bar{\alpha} }$ and $\frac{\partial\mathcal{L}}{\partial\beta } < \frac{\partial\mathcal{L}}{\partial \bar{\beta} }$ cannot hold almost surely (with probability one) for the optimal solution.

\textit{Case 5:} If we assume that $\frac{\partial\mathcal{L}}{\partial \alpha } < \frac{\partial\mathcal{L}}{\partial \bar{\alpha} }$ and $\frac{\partial\mathcal{L}}{\partial\beta } = \frac{\partial\mathcal{L}}{\partial \bar{\beta} }$ hold, we obtain $\alpha=0$. Similar to the reasoning given for Case 4,  we can conclude that the necessary condition for the optimality of this case holds with probability zero.

\textit{Case 6:} If we assume that $\frac{\partial\mathcal{L}}{\partial \alpha } > \frac{\partial\mathcal{L}}{\partial \bar{\alpha} }$ and $\frac{\partial\mathcal{L}}{\partial\beta } > \frac{\partial\mathcal{L}}{\partial \bar{\beta} }$ hold, we obtain $\alpha=1$ and $\beta=1$. Substituting these values in (\ref{StationaryPower3}), we obtain the necessary condition for $\frac{\partial\mathcal{L}}{\partial \alpha } > \frac{\partial\mathcal{L}}{\partial \bar{\alpha} }$ and $\frac{\partial\mathcal{L}}{\partial\beta } > \frac{\partial\mathcal{L}}{\partial \bar{\beta} }$~as 
\begin{IEEEeqnarray}{lll}\label{NecEquCase6}
    \sqrt{\frac{\gamma_2s_2}{\gamma_1s_1}} &\geq \frac{1-\mu}{\mu} \left[ 1+\left(\sqrt{\gamma_1s_1}+\sqrt{\gamma_2s_2}\right)^2 \right] \IEEEyesnumber \IEEEyessubnumber  \\ 
    \sqrt{\frac{\gamma_1s_1}{\gamma_2s_2}} &\geq \frac{1-\mu}{\mu} \left[ 1+\left(\sqrt{\gamma_1s_1}+\sqrt{\gamma_2s_2}\right)^2 \right], \IEEEyessubnumber 
\end{IEEEeqnarray}
respectively. The set of fading states $\mathbf{s}$ which satisfy both aforementioned conditions is non-empty only if the right hand sides of (\ref{NecEquCase6}a) and (\ref{NecEquCase6}b) are less than one which leads to $\mu>\frac{1}{2}$.

\textit{Case 7:} If we assume that $\frac{\partial\mathcal{L}}{\partial \alpha } = \frac{\partial\mathcal{L}}{\partial \bar{\alpha} }$ and $\frac{\partial\mathcal{L}}{\partial\beta } = \frac{\partial\mathcal{L}}{\partial \bar{\beta} }$ hold, we obtain $\alpha$ and $\beta$ from (\ref{StationaryPower3}) as 
\begin{IEEEeqnarray}{rll}
    \alpha &= \left(\mu-\frac{1}{2}\right)\frac{1+\gamma_1s_1+\gamma_2s_2}{\gamma_1s_1} \IEEEyesnumber \IEEEyessubnumber  \\ 
    \beta &= \left(\mu-\frac{1}{2}\right)\frac{1+\gamma_1s_1+\gamma_2s_2}{\gamma_2s_2}. \IEEEyessubnumber 
\end{IEEEeqnarray}
Moreover, we obtain the necessary condition for the optimality of this case as 
\begin{IEEEeqnarray}{rll}
    \min\{\gamma_1s_1,\gamma_2s_2\} \geq \left(\mu-\frac{1}{2}\right)\left(1+\gamma_1s_1+\gamma_2s_2 \right).
\end{IEEEeqnarray}

\textit{Case 8:} If we assume that $\frac{\partial\mathcal{L}}{\partial \alpha } = \frac{\partial\mathcal{L}}{\partial \bar{\alpha} }$ and $\frac{\partial\mathcal{L}}{\partial\beta } > \frac{\partial\mathcal{L}}{\partial \bar{\beta} }$ hold, we obtain $\beta=1$. Substituting this value in (\ref{StationaryPower3}), we obtain the optimal $\alpha$ as 
\begin{IEEEeqnarray}{rll}
    \alpha = \left[ \frac{-b_1+\sqrt{b_1^2+4a_1c_1}}{2a_1}  \right]^2,
\end{IEEEeqnarray}
$a_1=\frac{2-\mu}{\mu}\gamma_1S_1$, $b_1=\frac{1-\mu}{\mu}\sqrt{\frac{\gamma_1s_1}{\gamma_2s_2}}(1+\gamma_1s_1+\gamma_2s_2)$, and $c_1=1+\gamma_1s_1$. Moreover, the following conditions have  to hold for the optimality of this case
\begin{IEEEeqnarray}{rll}\label{CondApp1}
    \sqrt{\frac{\gamma_2s_2}{\gamma_1s_1}} &\leq \frac{1-\mu}{\mu} \left[ 1+\left(\sqrt{\gamma_1s_1}+\sqrt{\gamma_2s_2}\right)^2 \right]    \IEEEyesnumber \IEEEyessubnumber  \\ 
    \gamma_2s_2 &\leq \left(\mu-\frac{1}{2}\right)\left(1+\gamma_1s_1+\gamma_2s_2 \right), \IEEEyessubnumber 
\end{IEEEeqnarray}
where (\ref{CondApp1}a) and (\ref{CondApp1}b) are the necessary conditions for $\frac{\partial\mathcal{L}}{\partial \alpha } = \frac{\partial\mathcal{L}}{\partial \bar{\alpha} }$ to have a solution and for $\frac{\partial\mathcal{L}}{\partial\beta } > \frac{\partial\mathcal{L}}{\partial \bar{\beta} }$ to hold, respectively.

\textit{Case 9:} If we assume that $\frac{\partial\mathcal{L}}{\partial \alpha } > \frac{\partial\mathcal{L}}{\partial \bar{\alpha} }$ and $\frac{\partial\mathcal{L}}{\partial\beta } = \frac{\partial\mathcal{L}}{\partial \bar{\beta} }$ hold, we obtain $\alpha=1$. Substituting this value in (\ref{StationaryPower3}), we obtain the optimal $\beta$ as 
\begin{IEEEeqnarray}{rll}
    \beta = \left[ \frac{-b_2+\sqrt{b_2^2+4a_2c_2}}{2a_2}  \right]^2,
\end{IEEEeqnarray}
where $a_2=\frac{2-\mu}{\mu}\gamma_2S_2$, $b_2=\frac{1-\mu}{\mu}\sqrt{\frac{\gamma_2s_2}{\gamma_1s_1}}(1+\gamma_1s_1+\gamma_2s_2)$, and $c_2=1+\gamma_2s_2$. Moreover, the following conditions have  to hold for the optimality of this case
\begin{IEEEeqnarray}{rll}
        \sqrt{\frac{\gamma_1s_1}{\gamma_2s_2}} &\leq \frac{1-\mu}{\mu} \left[ 1+\left(\sqrt{\gamma_1s_1}+\sqrt{\gamma_2s_2}\right)^2 \right]     \IEEEyesnumber \IEEEyessubnumber  \\ 
    \gamma_1s_1 &\leq \left(\mu-\frac{1}{2}\right)\left(1+\gamma_1s_1+\gamma_2s_2 \right). \IEEEyessubnumber 
\end{IEEEeqnarray}

Note that the only necessary conditions for the optimality of Cases 6, 7, 8, and 9  can hold with a non-zero probability. Considering that these necessary conditions are mutually  exclusive,  see sets $\mathcal{K}_1$, $\mathcal{K}_2$, $\mathcal{K}_3$, and $\mathcal{K}_4$ in (\ref{MutualSet}) and Fig. \ref{FigPowerReg}, we obtain the optimal power sharing policy in (\ref{OptPower}).

\subsection{Optimal Lagrange Multiplier}

We note that given the optimal values of $q_k(\mathbf{s})$ and $[\alpha^{(k)}(\mathbf{s}),\beta^{(k)}(\mathbf{s})]$ in (\ref{OptPolicy}) and (\ref{OptPower}), respectively, all terms in  the constraint of the optimization problem in (\ref{ProbOpt}) can be calculated numerically for a given $\mu$, cf. (\ref{EqConst}). The optimal value of Lagrange multiplier $\mu$ is chosen such that the constraint in (\ref{ProbOpt}) is satisfied. Moreover, the optimal value of $\mu$ belongs to interval $(\frac{1}{2},1)$. To show this, we use  contradiction.

\textit{Case 1:} If $\mu\geq 1$, from (\ref{StationaryMode}), we obtain  $\frac{\partial\mathcal{L}}{\partial q_3(\mathbf{s}) } > \max \big\{\frac{\partial\mathcal{L}}{\partial q_1(\mathbf{s}) },\frac{\partial\mathcal{L}}{\partial q_2(\mathbf{s}) } \big\} $ which leads to $q_3(\mathbf{s})=1$, for $\forall \mathbf{s}$.  Moreover, from (\ref{StationaryPower3}), we obtain that $\frac{\partial\mathcal{L}}{\partial \alpha } \geq 0 \geq \frac{\partial\mathcal{L}}{\partial \bar{\alpha} }$ and $\frac{\partial\mathcal{L}}{\partial\beta } \geq 0 \geq \frac{\partial\mathcal{L}}{\partial \bar{\beta} }$ hold which leads to $\alpha=\beta=1$, i.e., only the cooperative messages are transmitted for mode $\mathcal{M}_3$. However, this is not possible since the SC-BSs have no cooperative messages to transmit as  the cooperative transmission modes $\mathcal{M}_1$ and $\mathcal{M}_2$ cannot be selected. Hence, $\mu\geq 1$ cannot hold for the optimal solution of  (\ref{ProbOpt}).

\textit{Case 2:} If $\mu\leq \frac{1}{2}$, from (\ref{StationaryPower3}), we obtain $\frac{\partial\mathcal{L}}{\partial q_1(\mathbf{s}) },\frac{\partial\mathcal{L}}{\partial q_2(\mathbf{s}) } > \frac{\partial\mathcal{L}}{\partial q_3(\mathbf{s}) } $. Hence, we obtain  $\alpha=\beta=0$, i.e., only the non-cooperative messages are transmitted for mode $\mathcal{M}_3$. However, this is not possible since the cooperative messages will never be transmitted to the MC-BS and be trapped in the buffers of both SC-BSs. This leads to the violation of the constraint in (\ref{ProbOpt}). Hence, $\mu\leq \frac{1}{2}$ cannot hold for the optimal solution of~(\ref{ProbOpt}).

To summarize, in this appendix, we have obtained the optimal mode selection and power sharing policies given in (\ref{OptPolicy}) and (\ref{OptPower}) in Theorem \ref{OptProt}, respectively. Moreover, the Lagrange multiplier $\mu$  satisfying the constraint in (\ref{ProbOpt}) is obtained, cf.  (\ref{EqConst}) in Theorem \ref{OptProt}, and its optimal value belongs to interval $(0,\frac{1}{2})$. This completes the proof.

}

\bibliographystyle{IEEEtran}
\bibliography{Ref_18_12_2016}
\end{document}